\documentclass[prl,twocolumn,notitlepage,superscriptaddress,longbibliography]{revtex4-1}
\usepackage{amsmath}
\usepackage{amssymb}

\usepackage[unicode=true,colorlinks=true,citecolor=blue,urlcolor=blue]{hyperref}

\usepackage{bm}
\usepackage{epsfig}

\usepackage[normalem]{ulem}

\renewcommand {\Im}{\mathop\mathrm{Im}\nolimits}
\newcommand{\GO}{\gamma_{\rm 1D}}
\renewcommand {\Re}{\mathop\mathrm{Re}\nolimits}

\renewcommand {\phi}{{\varphi}}
\newcommand {\rmi}{{\rm i}}

\newcommand {\sign}{\mathop{\mathrm{sign}}\nolimits}
\newcommand {\e}{{\rm e}}
\newcommand {\eps}{\varepsilon}

\begin{document}
\title{%
Bound state of distant photons in  waveguide quantum electrodynamics
}

\author{Alexander V. Poshakinskiy}
\affiliation{Ioffe Institute, St. Petersburg 194021, Russia}

\author{Alexander N. Poddubny}
\email{alexander.poddubnyy@weizmann.ac.il}
\affiliation{Department of Physics of Complex Systems, Weizmann Institute of Science, Rehovot 7610001, Israel}

\begin{abstract}
Quantum correlations between distant particles remain enigmatic since the birth of quantum mechanics. Here we predict a novel kind of bound quantum state in the simplest one-dimensional setup of two interacting particles in a box. Paradoxically, two entangled particles become localized at the opposite edges of the box even though their interactions at large distance should seemingly play no role. Such states could be realized in the waveguide quantum electrodynamics platform, where an array of superconducting qubits or cold atoms is coupled to a waveguide. We demonstrate how long-range waveguide-mediated couplings enable interaction-induced quantum states separated by large distances. Similarly to Majorana fermions in the Kitaev model, such bound state of distant photons is immune to short-range interactions and could find applications in robust quantum information processing. 
\end{abstract}
\date{\today}

\maketitle
\section{Introduction}

The quantum-mechanical problem of a particle mov-ing in one spatial dimension and confined in a box is probably the most paradigmatic model of quantum mechanics. A wave function of a single particle of mass $m$ in a box of size $N$ forms standing waves, $\psi(x)\propto \sin Kx$, where $K = \pi /N, 2\pi/N,...$ and the energy levels are given by $\hbar^2k^2/2m$. The situation becomes more interesting in the many-body case, when several quantum particles are put in the box and are allowed to interact with each other. For example, the problem of bosons with strong repulsive interaction can be solved exactly and the composite many-body wavefunction is proportional to a Slater determinant of wavefunctions of non-interacting particles, e.g., 
\[
\Psi(x_1,x_2)\propto {\sign (x_1-x_2)}[\psi_1(x_1)\psi_2(x_2)-\psi_2(x_1)\psi_1(x_2)]
\]
for a pair of particles. 
The strong repulsion of bosons thus effectively emulates the Pauli exclusion principle,  leading to a so-called fermionization~\cite{Lieb1963}. Another possibility is offered by bound many-particle states. The two attracting particles can form a bound pair so that their joint probability will decay with the characteristic length $a$. Such bound pair can propagate as a whole with the center-of-mass wave vector $K$ and can be quantized in a finite box, 
\[
\Psi(x_1,x_1)\propto \sin \left( K\,\frac{x_1+x_2}{2}\right)\e^{-|x_1-x_2|/a}\:.
\]
One of the instructive examples of such states is presented by a bound electron-hole pair in a semiconductor, an exciton, that is  confined in a quantum well~\cite{IvchenkoPikus}. If the well is wider than the exciton Bohr radius, the exciton is quantized as a whole. On the other hand, if the well width is smaller than the Bohr radius, the exciton is destroyed and electrons and holes are quantized independently, $\Psi(x_1,x_2)\propto \sin K_1x_1  \sin K_2x_2$ where $K_{1,2}$ are  the wave vectors of the corresponding particles.

In this work we consider  a novel kind of quantum state of two interacting particles in a box, different from those mentioned above, that we term as ``distant bound state", where the wave function has the form 
\begin{equation}
\Psi(x_1,x_2)\propto \e^{-x_1/a}\e^{-|N-x_2|/a}+(x_1\leftrightarrow x_2)\:.\label{eq:distant}
\end{equation}

Such state can be viewed as an entangled Bell state of particles, pinned by the interaction to the opposite sides of the box. The key ingredient, necessary to formation. of a state Eq.~\eqref{eq:distant}, is the strong interaction between the particles at large distance, that is necessary to repel them from each other. This requirement may seem very challenging and even self-contradictory, because the two particles, exponentially localized at the opposite edges of the structure should hardly interact. Any significant repulsion, pushing them to the opposite edges, does not seem feasible. However, we will demonstrate that this seeming paradox is resolved in the setup of waveguide quantum electrodynamics (WQED), where an array of natural or artificial atoms (such as superconducting qubits or quantum defects) is coupled to the waveguide~\cite{Roy2017,KimbleRMP2018,sheremet2021waveguide,Corzo2019,Oliver2022}. The WQED setup has built-in long-ranged interactions between the atoms, mediated by the photons propagating in the waveguide. Our goal is to demonstrate that distant bound states Eq.~\eqref{eq:distant} naturally arise in the WQED setup and that they are robust against fluctuating short-range interactions between the atoms. The distant bound state is a compound boson, but in a certain sense it reminds Majorana fermions in  the Kitaev's model, that were predicted to arise at the edges of a nanowire put upon a superconductor in a magnetic field~\cite{Oreg2010}. A pair of such Majorana states localized on the opposite edges of the wire forms a single ordinary fermionic excitation with zero energy. The latter appears to be partially  immune to dephasing because its two Majorana components cannot be mixed by a short-range perturbation, such as, e.g., Coulomb interaction~\cite{kitaev2001,wilczek2009}
In our setup, however, distant bound state emerges solely from atom-photon interactions, without any superconductivity or magnetic field. 

\section{Results}
\subsection{Qualitative origin of distant bound states}
\begin{figure}[bt]
\includegraphics[width=0.45\textwidth]{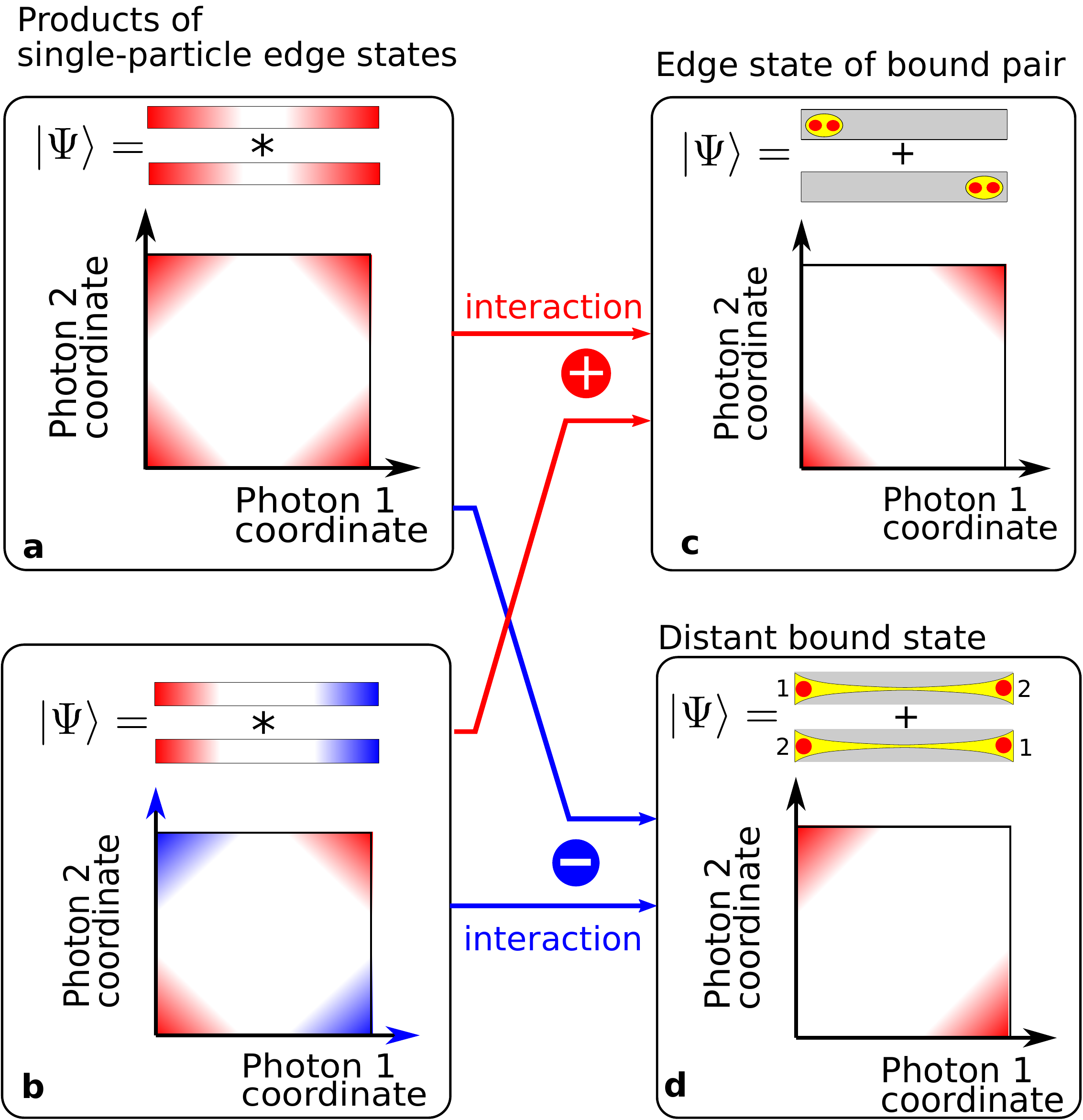}
\caption{{\bf Bound and edge  states.} 
Schematic illustration of two-photon wavefunctions for superpositions of even ($\bf a$)  and odd ($\bf b$) single-particle edge states and their interaction-induced combinations: edge state of bound pair ($\bf c$) and distant bound state ($\bf d$). 
}\label{fig:1}
\end{figure}
The main idea behind the origin of distant bound states is illustrated in Fig.~\ref{fig:1}.  Suppose that in the semi-infinite arrays there exist single-particle edge states, $|L\rangle$ and $|R\rangle$, localized at the left and right edges, respectively. Due to the mirror symmetry, the eigenstates of the finite array will be the even and odd combinations $|\psi_\pm\rangle=( |L\rangle \pm |R\rangle)/\sqrt{2}$. Now we  proceed to the double-excited states. Figures~\ref{fig:1}a and \ref{fig:1}b illustrate the product states with both particles either in the state $|\psi_+\rangle$ or $|\psi_-\rangle$, namely $|\Psi_\pm\rangle=|\psi^{(1)}_\pm\rangle |\psi^{(2)}_\pm\rangle$. In case of large structure, and when the interaction between the two particles is neglected, the states $|\Psi_\pm\rangle$ are degenerate. However, strong  interaction between the particles can mix these product states leading to the formation of new even and odd combinations $|\chi_\pm\rangle=(\Psi_+\rangle\pm |\Psi_-\rangle )/\sqrt{2}$, shown in Figs.~\ref{fig:1}b,\ref{fig:1}d. The even combination can also be interpreted as an edge state of the bound photon pair,
\begin{equation}
    |\chi_+\rangle=\frac{1}{\sqrt{2}}(|L_1\rangle|L_2\rangle+|R_1\rangle|R_2\rangle)
\end{equation}
where both particles are simultaneously localized either at the left or at the right edge of the structure (here the subscripts $1$, $2$ denote the particle numbers). In this work we focus on the odd combination 
\begin{equation}
    |\chi_-\rangle=\frac{1}{\sqrt{2}}(|L_1\rangle|R_2\rangle+|R_1\rangle|L_2\rangle)\:,
\end{equation}
which is an entangled Bell state of photons at the left and right edges of the array [Fig.~\ref{fig:1}(d)] and is equivalent to the distant bound state Eq.~\eqref{eq:distant}. 
\subsection{Numerical modeling}
\begin{figure*}[bt]
\centering\includegraphics[width=\textwidth]{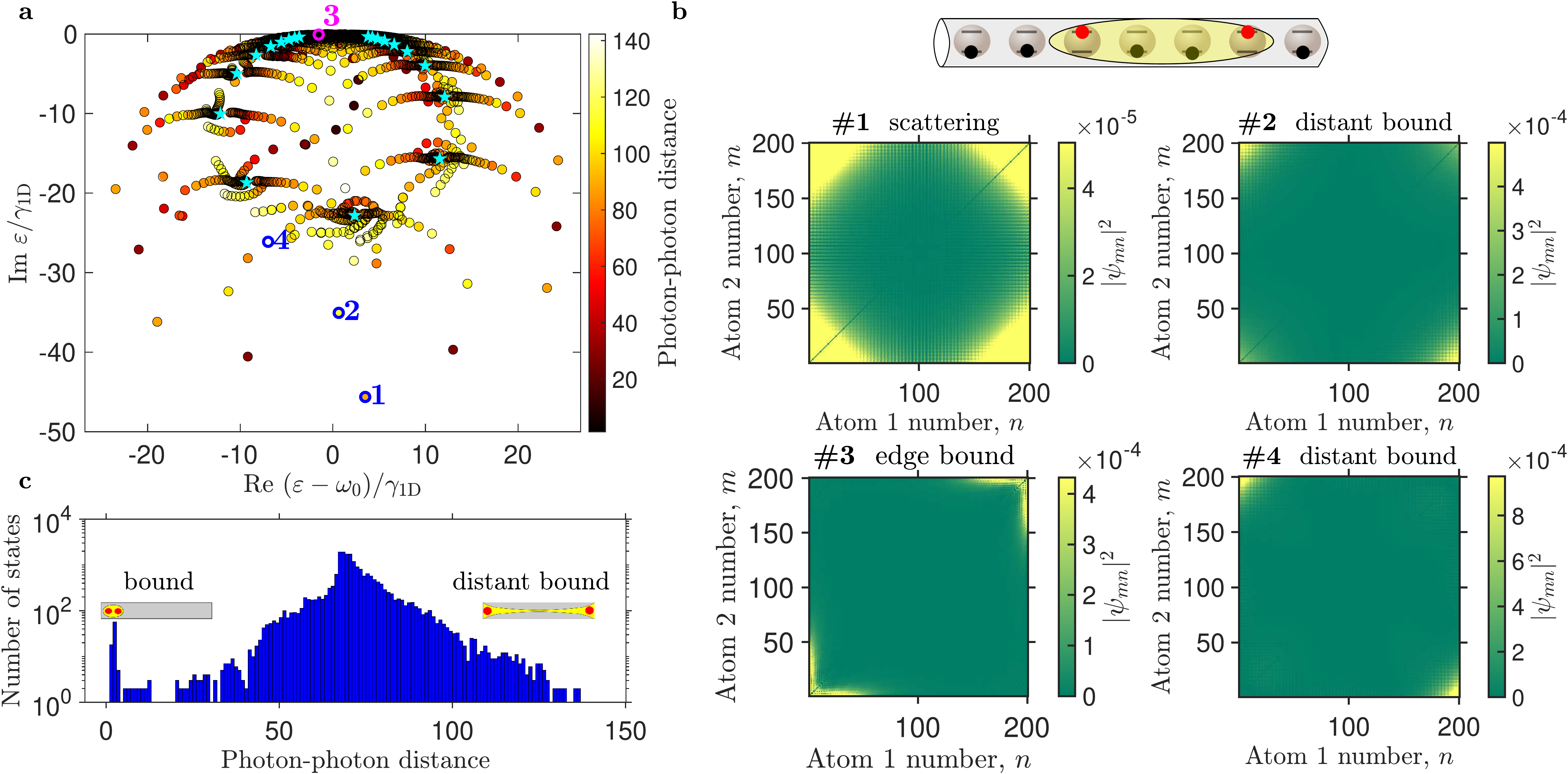}
\caption{{\bf Complex energy spectrum.} 
({\bf a}). Complex two-polariton energy spectrum of the array of $N = 200$ atoms. Color indicates the mean photon-photon distance $\rho$ calculated according to Eq.~\eqref{eq:rho}. Cyan stars show single-polariton energy spectrum. ({\bf b}).  Color maps show probability distribution $|\psi_{nm}|^2$ for four characteristic two-polariton states with the energies 
$(\eps-\omega_0)/\GO\approx 3.5-45.6\rmi, -7.0- 26.1\rmi,   11.1 -19.9\rmi, -1.6-0.1\rmi$, respectively, indicated with the corresponding numbers in (a). Top inset shows schematics of the array of atoms coupled to the waveguide with two interacting excitations.
(c) Histogram showing the distribution of the photon-photon distances $\rho$. 
 Calculation has been performed for  $N=200$ and $\varphi\equiv\omega_0d/c=1$.
 }\label{fig:2}
\end{figure*}

We consider an array of equidistant atoms coupled to a waveguide, schematically illustrated in the top inset in Fig.~\ref{fig:2}b. The system is described by the following effective Hamiltonian, written in the Markovian approximation~\cite{Caneva2015},
\begin{equation}
H=-\rmi \GO\sum\limits_{n,m}\sigma_n^\dag \sigma_m^{\vphantom{\dag}} \e^{\rmi \phi |m-n|}\:,\label{eq:H}
\end{equation}
where the energy is counted from the atomic resonance $\omega_0$ (we assume $\hbar=1$), $\sigma_n^\dag$ are the atomic raising operators, $\sigma_n^2\equiv 0$, $\varphi=\omega_0d/c$ is the phase gained by light travelling the distance $d$ between two neighboring atoms. The parameter $\GO\equiv \Gamma_{\rm 1D}/2$ is the radiative decay rate of  single atom into the waveguide. The key feature of the Hamiltonian Eq.~\eqref{eq:H} is the long-ranged waveguide-mediated coupling between the distant atoms.  We diagonalize the Hamiltonian Eq.~\eqref{eq:H} numerically in the domain of double-excited states $\sum_{n,m=1}^N\Psi_{nm}\sigma_n^\dag \sigma_m^\dag |0\rangle$ for finite $N$-atom arrays by solving the Schr\"odinger equation $H|\Psi\rangle=2\eps |\Psi\rangle$, see Ref.~\cite{sheremet2021waveguide} for the derivation details.

Figure~\ref{fig:2} presents our results obtained numerically for a finite array with $N=200$~atoms. The complex two-excitation energy spectrum is shown in Fig.~\ref{fig:2}a. Imaginary part of the eigenenergy $\eps$ describes the radiative losses into the waveguide. The points are colored according to the average photon-photon distance
\begin{equation}
    \rho=\sum\limits_{n,m=1}^N|n-m||\Psi_{nm}|^2\:\label{eq:rho}
\end{equation}
and we also show the histogram of the photon-photon distances distribution in Fig.~\ref{fig:2}c. This distribution has a broad peak at $\rho\approx 70\approx N/3$, corresponding to a pair of quasi-independent delocalized excitations. The tails of the distribution correspond to the bound photon pairs (small $\rho$) and distant bound states we focus on (large $\rho$). In order to provide more insight in the two-photon spectrum, we plot in 
Fig.~\ref{fig:2}b   the wavefunctions of four characteristic two-photon states of different types. For example, the state \#1 is a so-called scattering state, being a direct product of two symmetric combinations of left- and right-edge states, as shown in Fig.~\ref{fig:1}a. The state \#3 corresponds to both photons localized either close to the left- or to the right-edge of the array at the same time, and can be qualitatively understood as an edge state of bound photon pair [Fig.~\ref{fig:1}b]. Our  key observation in Figs.~\ref{fig:2}a is the existence of a large number of states, where the photon-photon distance $\rho$ is comparable  with the system size (yellow-colored points). The two-photon wave function for the most distant state \#2, with the largest value of $\rho$,  is shown in Fig.~\ref{fig:2}b. This state looks very similar to the Bell state of left- and right- edge photons, schematically illustrated in Fig.~\ref{fig:1}d. Another kind of distant bound state with a slightly different wavefunction is the state \#4. 
\begin{figure}[t]
\centering\includegraphics[width=0.47\textwidth]{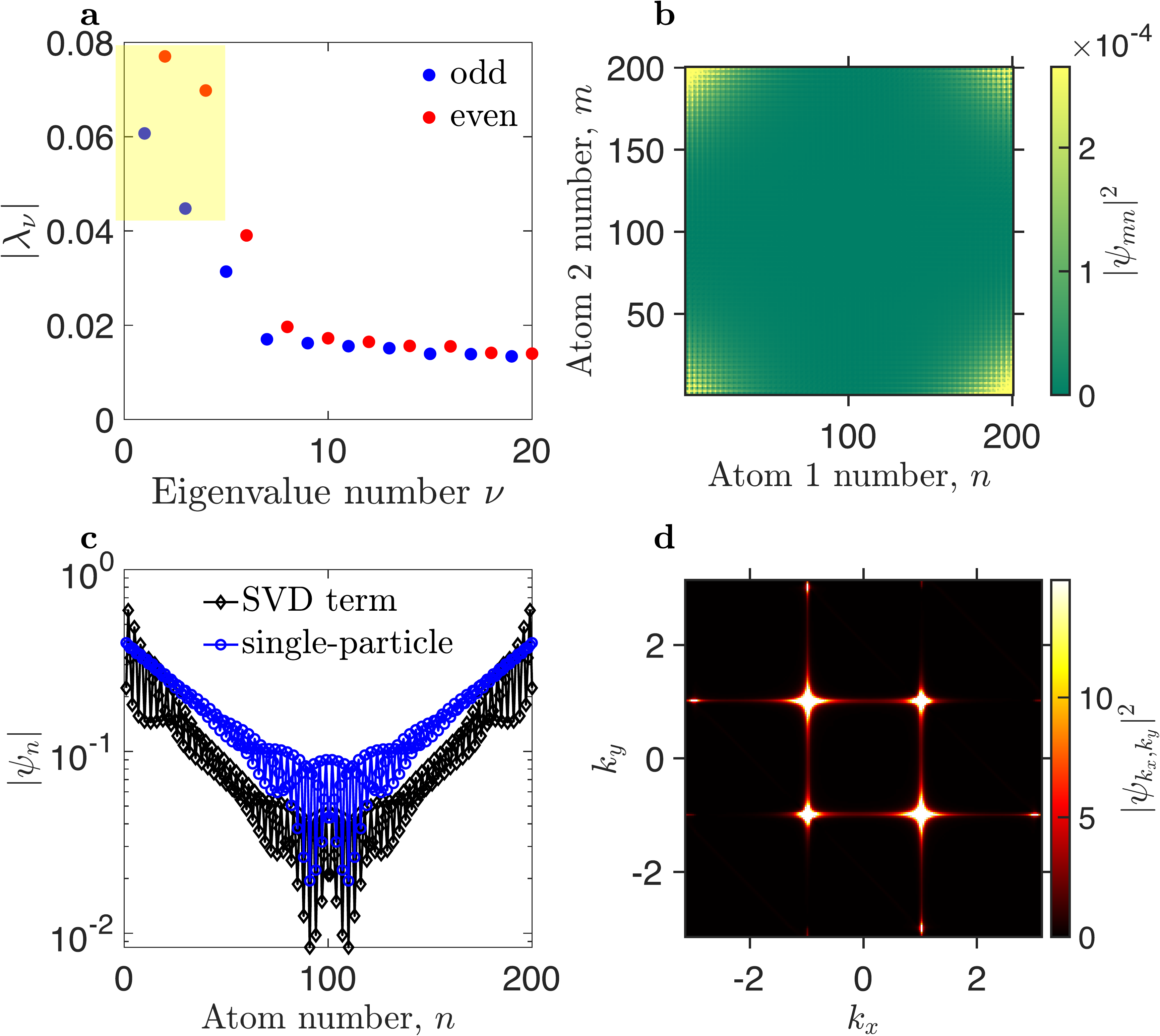}
\caption{{\bf Origin of distant bound states.}
({\bf a}) Largest singular values $\lambda$ entering the SVD expansion Eq.~\eqref{eq:svd} of the distant bound state \#2 from Fig.~\ref{fig:2}. Red and blue dots correspond to even and odd singular vectors, respectively. ({\bf b}) Approximated wavefunction of the distant bound state \#2 calculated including only four largest singular values in Eq.~\eqref{eq:svd}, indicated by the yellow rectangle in panel (a). ({\bf c}) Distribution of the wave function corresponding to the largest odd SVD term and the single-particle wavefunction of the brightest state. ({\bf d}) Color map of the two-dimensional Fourier transform of the wavefunction of the distant bound state \#2. 
}\label{fig:3}
\end{figure}

We now analyse the two-photon spectrum in more detail. Our goal is to understand the origin of single-particle edge states, and then to examine how their interaction leads to formation of two-particle bound states and finally demonstrate the robustness of the distant bound state against the disorder.
We start with the singular value decomposition of the distant bound state \#2 from Fig.~\ref{fig:2}b, 
\begin{equation}\label{eq:svd}
\Psi_{mn}=\sum\limits_{\nu=1}^N\lambda_\nu \psi_n^\nu\psi_m^\nu\:,
\end{equation}
where due to the bosonic symmetry $\Psi_{nm}=\Psi_{mn}$ left and right singular vectors coincide and can be chosen to satisfy the unconjugated orthogonality condition 
$\sum_{n=1}^m\psi_n^\nu \psi_n^\mu=\delta_{\mu\nu}$. The distribution of the 20 largest eigenvalues $\lambda_\nu$ is presented in Fig.~\ref{fig:3}a. Due to the mirror symmetry, the singular vectors $\psi_n^\mu$ are either odd or even and we denote the corresponding singular values by blue and red color, respectively. Figure~\ref{fig:2}b shows the approximation to the wavefunction, calculated leaving only four largest  singular values corresponding to  two odd and two even singular vectors in the expansion Eq.~\eqref{eq:svd}. Such two-term expansion  well approximates the exact wavefunction \#2 from Fig.~\ref{fig:2}b. The localization of the distant bound state at the edges of the array can be explained by the fact that the single-particle eigenstates are also localized, compare two corresponding wavefunctions, shown in Fig.~\ref{fig:2}c. These findings fully confirm our interpretation of the formation mechanism of the distant bound state illustrated in Fig.~\ref{fig:1}: it is formed due to the interaction-induced interference of the even and odd single-particle states localized at the opposite edges of the array.

\begin{figure}[t]
\centering\includegraphics[width=0.48\textwidth]{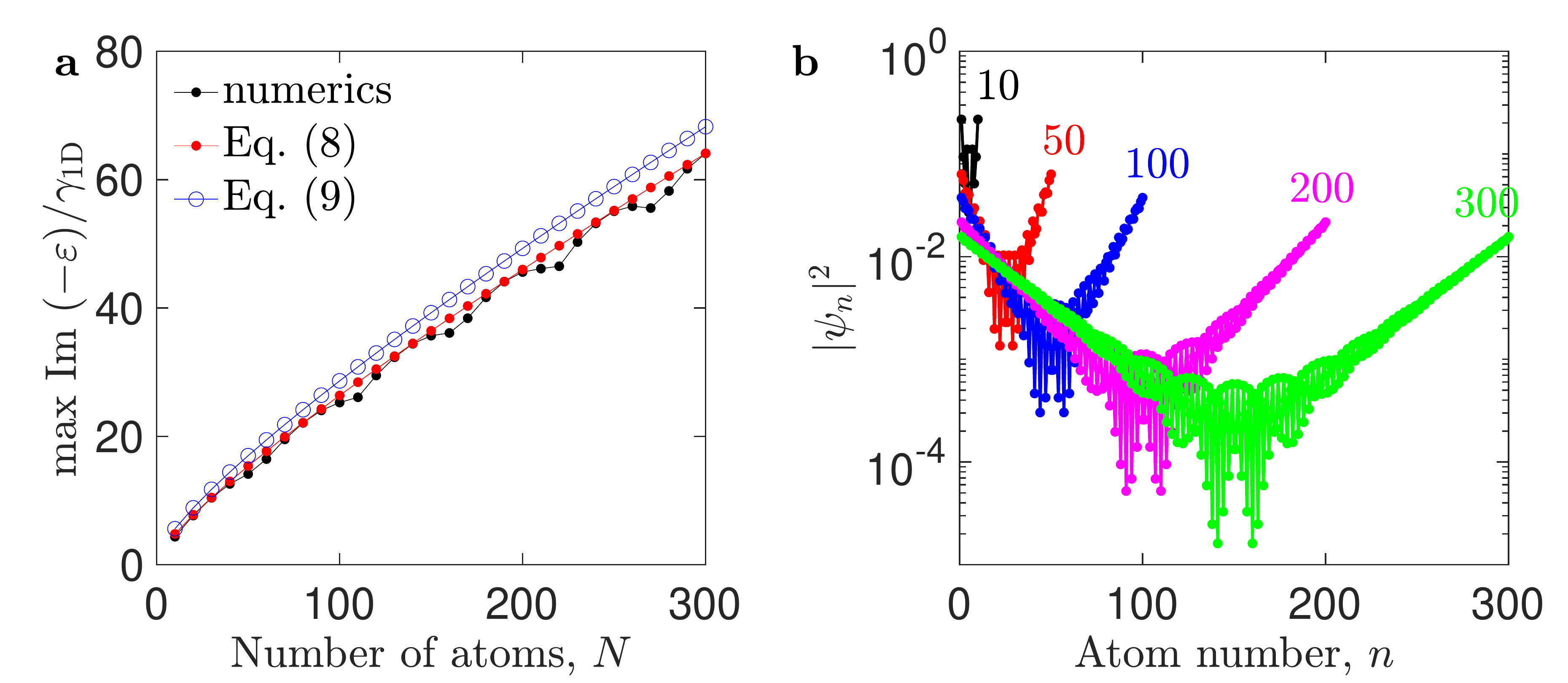}
\caption{
{\bf Scaling of single-particle eigenstates}. ({\bf a}) Dependence of the radiative decay rate of the brightest state on the number of atoms $N$ calculated numerically and analytically following Eqs.~\eqref{eq:lambert},\eqref{eq:lambert2}. ({\bf b}) Wave function of the brightest state depending on the length of the array. Other calculation parameters are the same as in Fig.~\ref{fig:2}.}\label{fig:4}
\end{figure}

It is also instructive to analyze the single-particle energy spectrum, obtained by solving the  Schr\"odinger equation $H|\psi \rangle=\eps |\psi$ with the ansatz $|\psi\rangle=\sum_{n=1}^N \psi_n \sigma^\dag |0\rangle$ and  shown in Fig.~\ref{fig:2}a by cyan stars.  Eigenmodes of the finite array can be presented as a superposition of two polaritonic Bloch waves \cite{Voronov2007JLu}:
\begin{equation}
\label{eq:eigpol}
\psi_n\propto \rho \e^{\rmi Kn}+\e^{-\rmi Kn},
\end{equation}
where $K$ is the polariton wave vector determined from the dispersion equation $\cos K=\cos\phi-\GO\sin\phi/\eps$ and $\rho=-(1-\e^{\rmi (\varphi-K)})/(1-\e^{\rmi (\varphi+K)})$
is the internal reflection coefficient of polaritons from the edge of the array .
Due to the radiative losses, the polariton eigenfrequencies $\eps$ in the finite array and the corresponding wave vector $K$ are complex. The states with the largest radiative decay rate have the real part of the  wave vector $\Re K(\varepsilon)$ close to $\pm \varphi$, which can be understood as a kind of phase synchronism condition facilitating photon emission~\cite{fedorovich2020disorder}. 
The phase synchronism is also evident from the Fourier transform of the two-photon wavefunction $\psi_{k_x,k_y}=\sum_{m,n=1}^N \e^{-\rmi k_x m-\rmi k_y n}\psi_{mn}$, shown in Fig.~\ref{fig:3}d that is concentrated near the points where $k_{x},k_y=\pm\varphi$.
The eigenfrequencies of even eigenmodes satisfy the analytical Fabry-Perot-like equation~\cite{fedorovich2020disorder,PoshakinskiyBorrmann} $\rho(\eps)\e^{\rmi K(\eps)(N+1)}=1$. Looking for the solution of this equation with 
$K\approx \varphi+\rmi \Im K$,  we obtaine the following approximate analytical equation for the decay rate of the brightest state,
\begin{equation}\label{eq:lambert}
    -\frac{\Im\eps}{\GO}\approx \frac{N}{W(2N\sin\varphi)}\:,
\end{equation}
where $W(x)$ is the Lambert $W$-function defined by the equation $W\e^W=x$. 
For large $N$ we use the approximation for the Lambert function, which yields
\begin{equation}\label{eq:lambert2}
    -\frac{\Im\eps}{\GO} \approx\frac{N}{\log(2N\sin\phi)-\log\log(2N\sin\phi)}\:.
\end{equation}
The dependence of the radiative decay on the array length, calculated numerically and analytically following Eqs.~\eqref{eq:lambert},\eqref{eq:lambert2} is shown in Fig.~\ref{fig:4}a. The decay rate increases with the number of atoms almost linearly due to the weak logarithmic growth of the Lambert-function and the corresponding eigenstate can be considered as a superradiant one.
Due to the radiative decay, the superradiant states have large imaginary part of the polariton wave vector $\Im K(\eps)>0$, and, according to Eq.~\eqref{eq:eigpol}, this leads to the decay of the wavefunction  from the edges towards the structure center, that is caused solely by the radiative losses.  The wavefunction of the brightest state calculated for arrays of different lengths is shown in Fig.~\ref{fig:4}b and indeed falls exponentially from the edges to the center.  An important observation from Fig.~\ref{fig:4}b is that the slope of the dependence $\ln |\psi_n|$ becomes smaller with the increase of the array length. The probability to find polariton in the center is $\sim N$ times smaller than in the bulk,  $|\psi_{N/2}/\psi_1|^2\propto 1/N$. This is in stark contrast to the conventional edge state where the ratio $|\psi_{N/2}/\psi_1|^2$ decays exponentially with the array length.  Mathematically, such scaling happens because the eigenfrequency of the superradiant state found from  Eq.~\eqref{eq:lambert} approximately satisfies the equation $\e^{-(N-1)\Im K(\eps)}\approx 1/(N\sin \varphi)$. 
 Such universal scaling is a direct manifestation of the non-Hermitian nature of the considered problem. For larger $N$ the radiative decay due to the photon escape through the edges becomes relatively less important and the localization length increases.



\begin{figure*}[t]
\centering\includegraphics[width=0.9\textwidth]{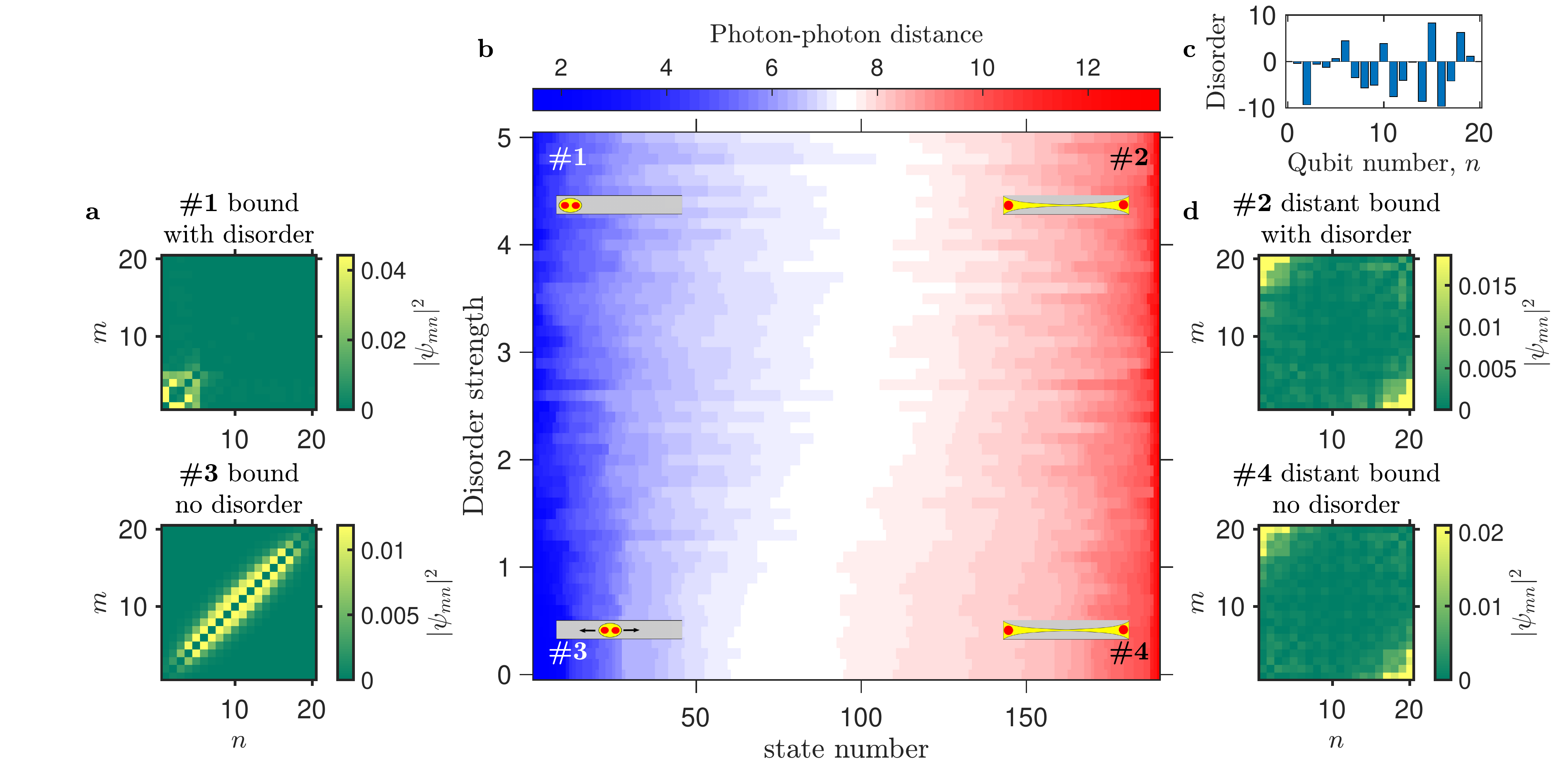}
\caption{{\bf Effect of disorder}. ({\bf a,d}) Wavefunctions of four characteristic two-photon states. ({\bf b}) Dependence of the photon-photon distances on the disorder strength $\chi$. Each horizontal line in the calculated 2D map corresponds to a single disorder realization. All the $N(N-1)/2$ eigenstates have been sorted according to the average photon-photon distance $p$ defined by Eq.~\eqref{eq:rho}. ({\bf c}) Distribution of the disorder amplitudes $\chi_n$ for the largest disorder strength $\chi = 5$. Calculation has been performed for $N=200$ and $\omega_0d/c = 1$. }\label{fig:disorder}
\end{figure*}
\subsection{Robustness against the disorder}
In order to investigate the stability of the distant bound states against the short-range disorder we add the following interaction term to the Hamiltonian Eq.~\eqref{eq:H}: 
\begin{equation}
V=\GO\sum\limits_{n=1}^{N-1}\chi_n (\sigma_n^\dag \sigma_n)(\sigma_{n+1}^\dag \sigma_{n+1})\:.\label{eq:V}
\end{equation}
Physically, such short-range interactions occur, e.g., due to van-der-Waals couplings between the atoms and they can be also implemented for superconducting qubits~\cite{Guimond2020}. Here, the coefficients $\chi_n$, characterizing the disorder strength, are independent random variables with zero mean value and the dispersion $\langle \chi_n^2\rangle \equiv  \chi^2 $, see Fig.~\ref{fig:disorder}c for a particular realization. The calculated dependence of the average photon-photon distance on the disorder strength is shown in Fig.~\ref{fig:disorder}. For vanishing disorder one can clearly distinguish the bound two-photon states with the small distance $p$ (state \#3 in Fig.~\ref{fig:disorder}a) and the distant bound states (state \#4 in Fig.~\ref{fig:disorder}b), see also the histogram Fig.~\ref{fig:2}c. For bound states two photons propagate together and their center-of-mass wave function forms a standing wave in the structure. These bound photon pairs are rather sensitive to the short-range disorder Eq.~\eqref{eq:V}. Increase of the disorder strength leads to the localization of the bound pairs, see the state \#1 in Fig.~\ref{fig:disorder}a. On the other hand, the distant bound states that we focus on are significantly less sensitive to the short-range interactions due to the increased photon-photon distance. This is evident from the persistence of the red-colored states with the large distance in the right edge of the diagram Fig.~\ref{fig:disorder}b. The most distant state is weakly affected by the disorder as can be seen by comparing the wavefunctions \#2 and \#4 in Fig.~\ref{fig:disorder}b. 
\section{Discussion}
\begin{figure}[t]
\centering\includegraphics[width=0.47\textwidth]{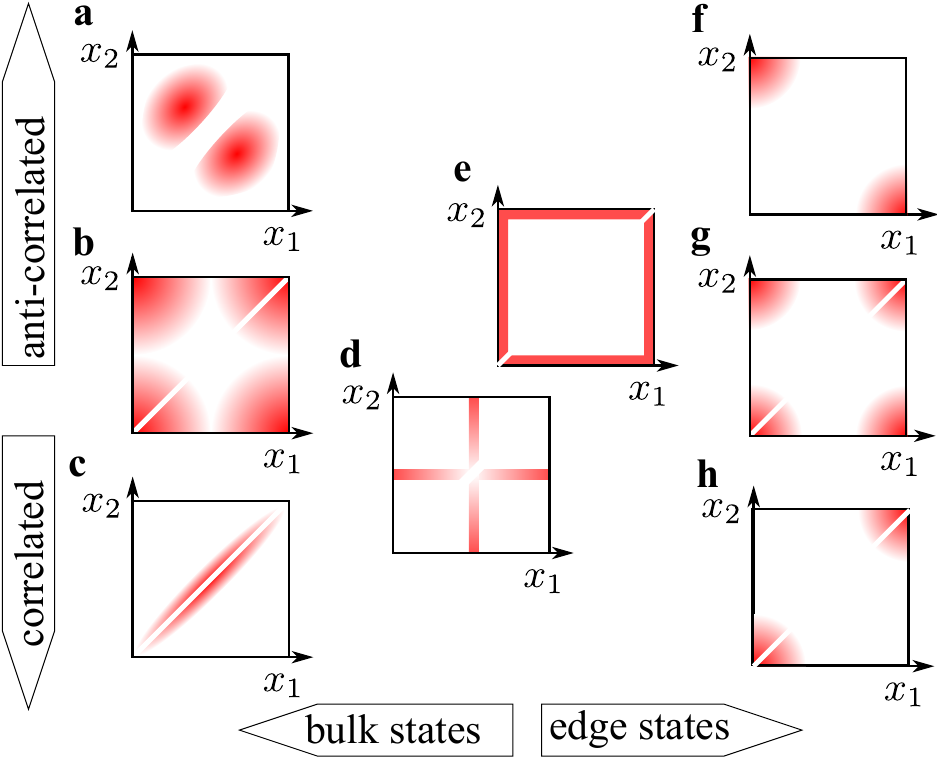}
\caption{{\bf Types of two-photon states. } The diagrams schematically illustrate two-photon wavefunction $|\Psi(x_1, x_2)|^2$ in a finite array of atoms  depending on the first and second photon coordinates. The states are grouped depending on whether the two photons are correlated or anticorrelated in space and located in the bulk of the structure or at its edges. Distant bound state, considered in this work, corresponds to  panel~(f). 
}\label{fig:6}
\end{figure}

In this section we try to present a bird-eye view on the two-photon quantum states in the finite array of atoms coupled to a waveguide. Spatial false-color maps of two- photon joint probabilities $|\Psi(x_1, x_2)|^2$ are presented in the schematic diagram Fig.~\ref{fig:6}. They are grouped depend- ing on relative distance between the two photons (larger distance corresponds to the upper panels) and depending on whether the photons are located mostly at the structure edges (right panels) or in the bulk (left panels). The bulk states  are most simple to understand. They are limited to the scattering states, where two photons are quasi-independent (Fig.~\ref{fig:6}b), fermionized states with increased radiative lifetime (Fig.~\ref{fig:6}a)~\cite{Molmer2019}, where the average photon-photon distance is increased and bound photon pairs (Fig.~\ref{fig:6}c)~\cite{Zhang2020PRR,poddubny2020quasiflat}. To the best of our knowledge, neither fermionized states nor bound states have not been directly observed in experiments yet. However, the bound states are well known in other setups. They have been observed for cold atoms in optical lattices~\cite{Winkler2006}, correlated two-photon quantum walks have recently been experimentally studied for superconducting quantum processors~\cite{Yan2019} and even three- photon bound states were seen for light interacting with Rydberg atomic states~\cite{Liang2018,drori2023}. Moreover, tunable photon bunching and antibunching, recently realized in the WQED setup with cold atoms coupled to the nanofiber, is in fact mediated by the two-photon bound states~\cite{Prasad2020}. Thus, direct experimental observation of the two-photon bound states does not seem to be in the realm of impossible for state-of-the art setups. The most advantageous platform should probably be based on superconducting qubits, since it allows access (excitation and probing) of individual qubits~\cite{zanner2021coherent}. As demonstrated by the calculation in Fig.~\ref{fig:disorder}, the array of 20 qubits should already be sufficient to observe the bound states as well as other quantum states discussed below. 
Figures~\ref{fig:6}d,~\ref{fig:6}e present unusual types of two-photon quantum states, predicted in our previous works~\cite{Zhong2020,Poshakinskiy2020,Poshakinskiy2021Chaos}. They manifest interaction-induced localization when one of the two indistinguishable photons forms a standing wave that induces a trapping potential for the other photon. This second photon can be localized either in the center (Fig.~\ref{fig:6}d) or at the edge (Fig.~\ref{fig:6}e) of the array. Since standing wave is delocalized, such states can be considered as lying in between bulk and edge states, for example for the state in Fig.~\ref{fig:6}e one of the photons is always in the bulk while another one is always at the edge. The two photons can be both correlated and anticorrelated in space depending on whether the trapping is in the anti-node or in the node of the standing wave. Thus, while being quite interesting from the fundamental side,
the states in Figs.~\ref{fig:6}d, ~\ref{fig:6}e do not seem optimal to enhance or suppress the photon-photon spatial correlations. 
Finally, we proceed to the states in Figs.~\ref{fig:6}f,g,h, where both photons are at the edges of the array. Depending on the specifics of two-photon interactions, the photon pair can be quasi-independent (Fig.~\ref{fig:6}g), correlated (Fig.~\ref{fig:6}h) or anti-correlated (Fig.~\ref{fig:6}f). As discussed above, the latter state, put forward in this work, is the most robust against short-range interactions between the photons. 

To summarize, waveguide quantum electrodynamics, that has become a separate research field only relatively recently, is a very promising platform to control two-photon correlations~\cite{Prasad2020}. Even the simplest two-body problem in WQED manifests a number of quite unusual two-photon states. It is not clear whether our classification in Fig.~\ref{fig:6} is complete and how it can be extended for larger number of particles~\cite{Zhong2021three} or even in the many-body regime~\cite{fayard2021manybody} but we can expect beautiful fundamental phenomena that will be also hopefully soon complemented by experimental demonstrations and even practical applications for the emerging quantum industry. 

%


\end{document}